\documentclass[runningheads]{llncs}
\usepackage{color}
\usepackage{amssymb}
\usepackage{graphicx}
\usepackage{framed}
\usepackage{amsmath}
\usepackage{url}
\usepackage{multicol}
\usepackage{listings}
\usepackage{xcolor}
\usepackage{array}
\usepackage{float}
\usepackage{graphicx}
\usepackage{subcaption}
\usepackage{booktabs}
\usepackage{tabularx}

\usepackage{rotating}

\restylefloat{table}
\newcolumntype{P}[1]{>{\centering\arraybackslash}p{#1}}

\colorlet{punct}{red!60!black}
\definecolor{background}{HTML}{EEEEEE}
\definecolor{delim}{RGB}{20,105,176}
\colorlet{numb}{magenta!60!black}
\lstdefinelanguage{json}{
	basicstyle=\normalfont\ttfamily,
	numbers=left,
	numberstyle=\scriptsize,
	stepnumber=1,
	numbersep=8pt,
	showstringspaces=false,
	breaklines=true,
	frame=lines,
	backgroundcolor=\color{background},
	literate=
	*{0}{{{\color{numb}0}}}{1}
	{1}{{{\color{numb}1}}}{1}
	{2}{{{\color{numb}2}}}{1}
	{3}{{{\color{numb}3}}}{1}
	{4}{{{\color{numb}4}}}{1}
	{5}{{{\color{numb}5}}}{1}
	{6}{{{\color{numb}6}}}{1}
	{7}{{{\color{numb}7}}}{1}
	{8}{{{\color{numb}8}}}{1}
	{9}{{{\color{numb}9}}}{1}
	{:}{{{\color{punct}{:}}}}{1}
	{,}{{{\color{punct}{,}}}}{1}
	{\{}{{{\color{delim}{\{}}}}{1}
	{\}}{{{\color{delim}{\}}}}}{1}
	{[}{{{\color{delim}{[}}}}{1}
	{]}{{{\color{delim}{]}}}}{1},
}

\urldef{\mailsa}\path|firstname.lastname@Anonymised.org|

\begin{document}


\title{Characteristics of Dataset Retrieval Sessions: Experiences from a Real-life Digital Library}
\titlerunning{Characteristics of Dataset Retrieval Sessions}
%
%
\author{Zeljko Carevic, Dwaipayan Roy, Philipp Mayr}
\authorrunning{Z. Carevic et al.}
%
\institute{GESIS –- Leibniz Institute for the Social Sciences, Cologne, Germany}
\maketitle              
%

\begin{abstract}		
Secondary analysis or the reuse of existing survey data is a common practice among social scientists. Searching for relevant datasets in Digital Libraries is a somehow unfamiliar behaviour for this community. 
Dataset retrieval, especially in the social sciences, incorporates additional material such as codebooks, questionnaires, raw data files and more. Our assumption is that due to the diverse nature of datasets, document retrieval models often do not work as efficiently for retrieving datasets. One way of enhancing these types of searches is to incorporate the users' interaction context in order to personalise dataset retrieval sessions. As a first step towards this long term goal, we study characteristics of dataset retrieval sessions from a real-life Digital Library for the social sciences that incorporates both: research data and publications.
Previous studies reported a way of discerning queries between \textit{document search} and \textit{dataset search} by query length.   In this paper, we argue the claim and report our findings of an indistinguishability of queries, whether aiming for a dataset or a document.  Amongst others, we report our findings of dataset retrieval sessions with respect to query characteristics, interaction sequences and topical drift within 65,000 unique sessions. 
\end{abstract}

\section{Introduction}
With the vast availability of research data on the Web within the Open Data initiatives, searching for it becomes an increasingly important and timely topic. The Web hosts a whole range of new data species, published in structured, unstructured and semi-structured formats -- from web tables to open government data portals, knowledge bases such as Wikidata and scientific data repositories. This data fuels many novel applications, for example, fact checkers and question answering systems, and enables advances in machine learning, artificial intelligence and information retrieval.

Dataset retrieval has emerged as an independent field of study from the text retrieval domain. The latter is well-known in information retrieval (IR) with research leading to significant improvements. Dataset retrieval, on the other hand, represents a challenging sub-discipline of information retrieval with substantial differences in comparison to traditional document retrieval~\cite{chapman2020survey,koesten2018}. Datasets, especially in disciplines such as the social sciences, often encompass complex additional material such as codebooks (incl. variable descriptions), questionnaires, raw data files and more. Due to the higher complexity of datasets, the applicability of IR models build mainly for document retrieval is questionable. In addition, the motivations and information needs of researchers seeking for datasets are too manifold to be supported by out-of-the-box retrieval technologies. Disciplines that encourage the re-use of datasets or secondary analysis, such as, the social sciences might thus not be supported sufficiently during dataset retrieval. One way of supporting users during dataset retrieval is the development of an integrated dataset retrieval system that employs advances from established document retrieval systems and adopts these techniques to the field of dataset retrieval. 
Our long term goal in the project ConData\footnote{\url{http://bit.ly/Condata}} is to develop an effective dataset retrieval system, that incorporates personalised searching by employing contextualised ranking features which aim at tailoring search results towards the users' information needs.
In order to develop a contextualised dataset retrieval approach, it is necessary to first gain a better understanding of different characteristics during dataset retrieval. Obtaining these kinds of behavioural data is usually hard. We address this shortcoming by analysing real-life user behaviour within a Digital Library for research data and related information for the social sciences~\cite{hienert2019digital}. As an initial outcome of this study, we report our findings on comparing dataset retrieval with document retrieval sessions corresponding to query characteristics, interaction sequences and topical drift within 65,000 unique search sessions. 
\vspace{-2mm}

\section{Related Work and Motivation}\label{sec:rel-work}
\vspace{-2mm}

Although started as a fundamental database task, the diverse nature of searched entities (which can be images, graphs, tables etc.) establishes dataset retrieval as a research domain for itself.
The distinctive aspects 
of dataset retrieval regarding complex information needs (and in turn, query formulations) make it a difficult process in comparison to document search~\cite{cafa2011struct,kern2015diff,koesten2018}. However, traditional keyword-based retrieval approaches are still in use in the domain of dataset retrieval although they are observed to be less effective for the task~\cite{chapman2020survey}. In order to exploit the additional information available for datasets, researches have been going on~\cite{googledataset,chen-CIKM2019} to achieve further improvement. 

An important sub-task during a retrieval session is to characterise the query to understand whether the search intent is of document or dataset. 
Considering the diversity in nature between dataset retrieval and document retrieval, an integrated search system (having both, datasets as well as documents as a repository) would benefit in selecting appropriate searching mechanism if the query intent  is recognized.
However, in~\cite{kacprzak-WWW2018}, Kacprzak et al. reported the difficulty in understanding the users' intent when performing dataset search.
They have subtly drawn a co-relation between query length and the type of query, and concluded with a suggestion to use longer queries for dataset retrieval.
Experimented in an artificial setting without a naturalistic information need, however, they concluded that their observation could be considered as an approximation of the user behaviour for comparing dataset and document search.

Few of the works on studying user behaviour in dataset search have been done examining queries submitted to open data portals and online communities~\cite{chen-CIKM2019,kacprzak-WWW2018}. However, in~\cite{jansen2006search}, Jansen and Spink concluded that it is not possible to directly compare the results of a transaction log analysis across different search engines.

In this work, we focus on characterising the users' intent when performing publication (document) search and research data (dataset) search\footnote{The words (\emph{document}, \emph{publication}) and (\emph{dataset}, \emph{research data}) are used interchangeably in the rest of the paper to imply the same concept.}.

\vspace{-3mm} 
\section{Experimental Materials}\label{sec:use}
\vspace{-2mm}
We conduct our experiment in a real-life Digital Library for the social sciences\footnote{Accessible via: \url{https://search.gesis.org}. See details in \cite{hienert2019digital}.}. This integrated search system (ISS) allows users to search across different data collections: research datasets, publications, survey variables, questions from questionnaires, survey instruments and tools for creating surveys. The focus of the following study is on datasets and publications.  The collection covering research datasets comprises 6,267 studies that are collected within our institution and 107,595 studies coming from other institutes. 
The collection covering publications comprises 48,234 records mainly as open accessible articles from the social sciences. Information items are interlinked whenever possible to allow a better findability and reuse of the data. 
The ISS uses category facets which enable a user to switch between data types. Furthermore, category facets ensure that result lists contain exactly one data type at a time. 
The ISS is mainly used by social scientists. A thorough report about the technical system, the content and its users can be found in  \cite{hienert2019digital}.

The user interactions within the ISS are anonymously logged, which makes it possible to study user behaviour on a larger scale. Amongst others, the log covers user actions such as \emph{queries submitted}, \emph{record views}, \emph{browse/filter operations}. For this study, we considered all search sessions from January 2018 to December 2019.
Sessions and their corresponding identifiers are not bound to a timeout. Instead, a session expires in ISS on termination of the Web browser. In order to determine a realistic session timeout, we decided to consider sessions exceeding an inactivity of 30 minutes as a new session. After this operation, we identified 30,695 dataset retrieval sessions and 34,550 sessions that were focused on publications.

Given a query \texttt{Q}, ISS returns a list of distinct categories such as \textbf{``research data'', ``publications''} along with \emph{``variables \& questions'', ``instruments \& tools''}  from which a user can choose to retrieve a corresponding result set. For this study, we are interested in those sessions containing queries that led to record views either in the category \emph{research data} or in \emph{publication}. 
We discriminate the \emph{research data search} and \emph{publication search} from the log based on the type of the succeeding record viewed by the user. We categorise a query as a publication search (or, dataset search) if the user has viewed a record of type publication (or research data) immediately after submitting the query to ISS. 
Finally, we extract only those sessions that are either of type \emph{research data} or \emph{publication}. In total, our preprocessed log file consists of 142,028 rows. The rows in the log represent queries submitted by users (identified by a session fingerprint) and corresponding record views which are either of type publication or research data. The former type accounts for 79,931 records and the later for 62,097 records. 
Certain preprocessing steps are necessary before analysing the transaction log: we remove sessions having queries that are either empty or contain unrecognisable characters (which might result from erroneous encoding).
\vspace{-3mm}
\section{Results and Observations}
\vspace{-3mm}
In this section, we present the results of our transaction log analysis. 
First, we summarize the results of our query characterisation in Section~\ref{subsec:query-charac}.
We compare and contrast dataset search and publication search on the basis of session-level information and sequential interaction information, respectively in Section~\ref{subsec:session} and Section~\ref{subsec:interact}.
\vspace{-3mm}
\subsection{Query Characterization}\label{subsec:query-charac}
In this study, we try to differentiate queries on the basis of their search intent (publication or research data search). In Table~\ref{tab:query-stat}, we present the basic statistics of queries with respect to publications and datasets.
\begin{table} [h]
	\centering
	\caption{
	Average statistics comparing queries for dataset and publication search. 
	}
	\begin{tabular}{ l r r } \toprule
		& Datasets & Publications \\ \midrule
		Total query count & 62,097 & 79,931\\
		Unique query count & 18,706 ($30.12\%$) & 33,228 ($41.57$\%) \\
		Avg. query length (char) & 15.93 & 19.67 \\
		Avg. query length (terms) & 1.89 & 2.07\\ 
		Queries with digits (\%) & 21.57\% & 3.22\%\\
		\bottomrule
	\end{tabular}
	\label{tab:query-stat}
\end{table}
\vspace{-2em}

The following observations can be drawn from Table~\ref{tab:query-stat}.
\begin{itemize}
    \item Publication search is more common than dataset search, with almost 28\% more submitted queries, in the ISS. This is in line with the observations already made in \cite{hienert2019digital}.
    \item Dataset search queries are much more repetitive than publication search queries with 69.88\% queries getting re-issued to the search system; in contrast, the queries are less repeated (58.43\%) for publication searches.
    We can interpret this observation by the variety of forms in representing the information need for publication searches (as compared to dataset search). 
    \item On average, the length of a dataset search query (measured by the number of characters\footnote{Character count is used considering the linguistics of German language; the queries submitted to the ISS are mixed, some in German and others in English.} as well as the number of terms in the query) is less as compared to publication search.
    This observation is in conflict with the notion presented in~\cite{kacprzak-WWW2018}, where the authors suggested issuing longer queries for dataset search.
    The reason can be a difference in the experimental settings of our study and~\cite{kacprzak-WWW2018} where the authors acknowledge the artificial, crowd-sourced nature of their study.
    \item Queries for dataset search significantly more often contain numerical digits as compared to queries for publication search. Research data includes a significant number of periodic records which are titled mentioning the periods (e.g. \texttt{allbus 2014, allbus 2016} etc. which refer to a biennial survey conducted since 1980).
\end{itemize}
\vspace{-4mm}
\subsection{Analyzing Sessions} \label{subsec:session}
In Table~\ref{tab:search-session}, we report the average number of record views for dataset search and publication search in a session.
From the table, we can see that the number of record views per session is higher for publication search than for dataset search.
This implies that users having a publication search intent are expected to view more items than for a dataset search intent. In other words, we assume that the information need for a dataset search can be addressed by a comparatively less number of record views than publication search.
\vspace{-2em}
\begin{table} [h]
	\centering
	\caption{Number of record views per session with different search intent.   }
	\begin{tabular}{ l c c } \toprule
		& Datasets & Publications\\ \midrule
        Avg. record views per session &2.02 & 2.31 \\
        Avg. record views per session (unique) & 1.61 & 2.06 \\
		\bottomrule
	\end{tabular}
	\label{tab:search-session}
\end{table} 
\vspace{-3em}
\subsection*{Session Diversity}\label{diversity}
\vspace{0px}
In a single session, a user could have multiple information needs and might have issued multiple queries to ISS.
In order to identify the diversity of the information need, an elementary way would be to observe the similarity of the issued queries.
However, being keyword queries, term overlap based similarity measurements, like IR-based TF-IDF model or a set-based Jaccard similarity model, would perform poorly when computing similarities among queries.

To have a better understanding of the diversity in information needs, an appropriate approach would be to inspect the similarity of viewed records: \emph{intra-record similarity is inversely proportional to the underlying diversity of a session}~\cite{angel2011}. 
We hypothesise that a heterogeneous set of viewed records indicates high diversity. 

In a single session $S$, let a user has viewed a set of records $\{r_f,\cdots, r_l\}$ ($r_i\in \{\text{publication}, \text{research data}\}$).
To determine whether a session can be considered as homogeneous, we measure the similarity between the first ($r_f$) and last ($r_l$) encountered record.
In order to do this, however, a similarity threshold value is needed to be fixed with annotated training data.
Instead, we apply the \emph{More Like This (MLT)}\footnote{{\url{https://bit.ly/MLT-elastic}}} module that is readily available in Elasticsearch. In the MLT module, similarity is computed using BM25 similarity between a given document (\emph{seed} document) and all the documents in the collection; it returns a list of documents which are similar in content with the seed document.
This approach, in comparison to query similarity, enables us to utilise the set of descriptive metadata to determine the similarity between documents while at the same time, being more robust to query modifications. 

For a session $S$, we define a tuple $(r_f, r_l)$ consisting of the first and last viewed record.
We consider $r_f$ to be the seed document for the \emph{MLT} module. 
For both, publications and datasets we retrieve top $k$ similar items for the seed ($r_f$) using \emph{MLT} module. 
If the last viewed record $r_l$ is present within the top $k$ more-like-this records, we consider the session as topically homogeneous.
However, choosing an appropriate $k$ is crucial in understanding the diversity of the session.
For this study, we experiment with setting $k$ to $100$ for a lenient understanding, and to $5$ for a more rigorous and restricted understanding of diversity.

\begin{figure}[t]
    \centering
    \includegraphics[width=1\linewidth]{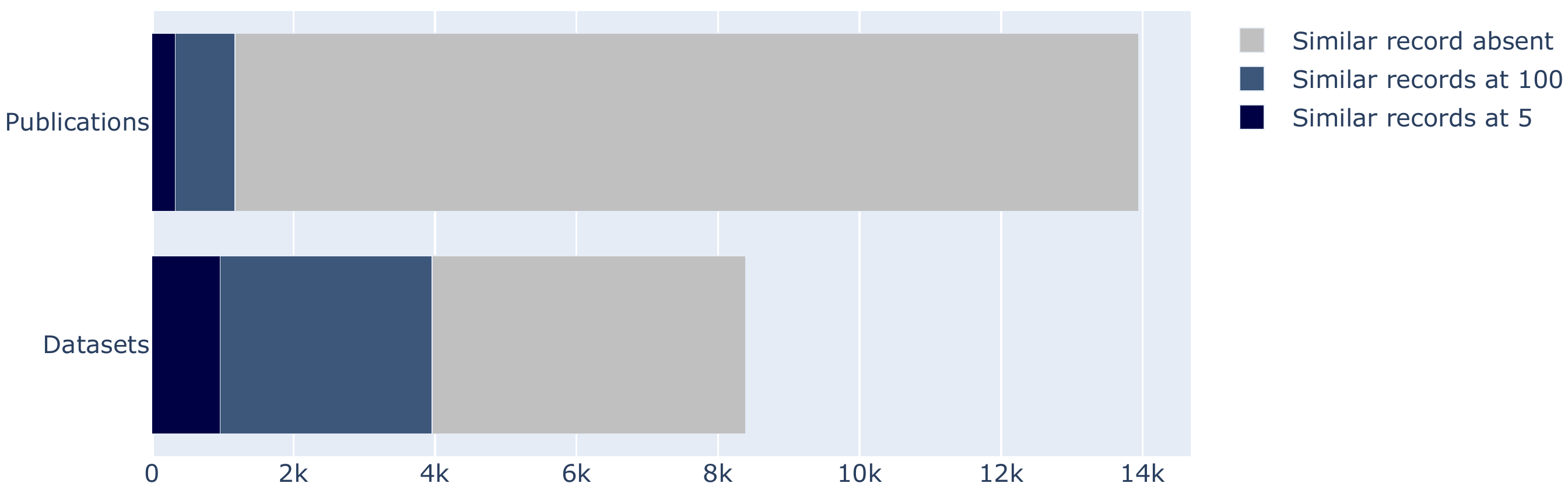}
    \caption{Session diversity at the top 100 and the top 5 similar records.}
    \label{fig:diversity}
    \vspace{10px}
\end{figure} 

The result of this analysis is presented graphically in Figure~\ref{fig:diversity}.
In the figure, the light grey shade corresponds to sessions for which the last record $r_l$ is not found within the top $100$ more-like-this records.
The blue and dark blue shades indicate the number of sessions for which the last record $r_l$ has been located respectively within the top $100$ and the top $5$ records as returned by \emph{MLT} module.  Note that this analysis is not applicable to those sessions having only one record view.

For dataset search (presented at the bottom of Figure~\ref{fig:diversity}), we note that approximately 11\% sessions (particularly $964$) are seen to be very focused to a particular topic (dark blue) for which the last viewed record ($r_l$) has been found within the top $5$ more-like-this items.
The last record is found within top $100$ more-like-this items for $2993$ sessions (blue) which accounts for $35.7\%$.
However, the publication search sessions seem much more diverse: we found only $846$ sessions ($6.1\%$) having a last record contained in the top 100 more-like-this items and only 329 sessions ($2.3\%$) for which the last record has been found within top $5$ \emph{MLT} entries. 

The topical diversity and homogeneity of a session for publication and dataset search is even more evident when we consider the similarity scores provided by the \emph{MLT} module ($\texttt{sim-score}(r_f, r_l)>0$). On average, a dataset search retrieved a similarity score of $\text{Top} 100 = 325.02, \text{Top} 5 = 475.0$ while the similarity score for publications  was only $\text{Top} 100 = 70.8$ and  $\text{Top} 5 = 117.9$. 
From this analysis, we can conclude that dataset retrieval sessions are much more focused than publication search sessions, and the searched datasets in a single session are more densely coupled than the searched publications.
\vspace{-2mm}
\subsection{Interaction Sequences}\label{subsec:interact}
\begin{figure}[h]
\begin{subfigure}[b]{0.5\textwidth}
   \includegraphics[width=1\linewidth]{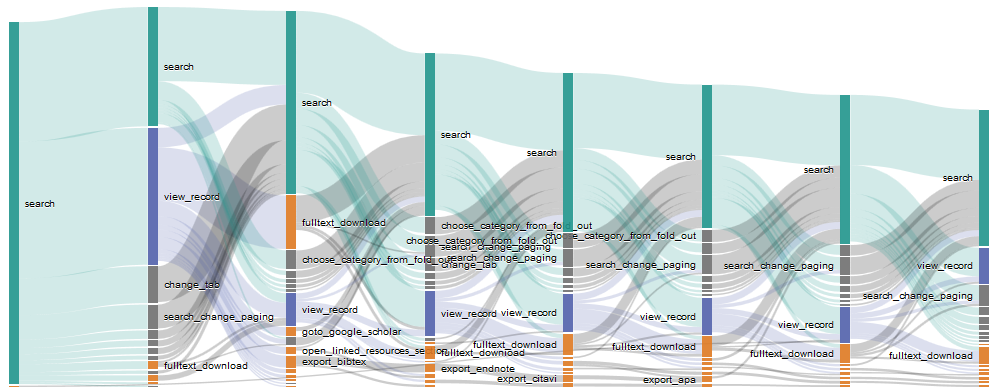}
   \caption{Publications}
   \label{fig:Ng1} 
\end{subfigure}
\begin{subfigure}[b]{0.5\textwidth}
   \includegraphics[width=1\linewidth]{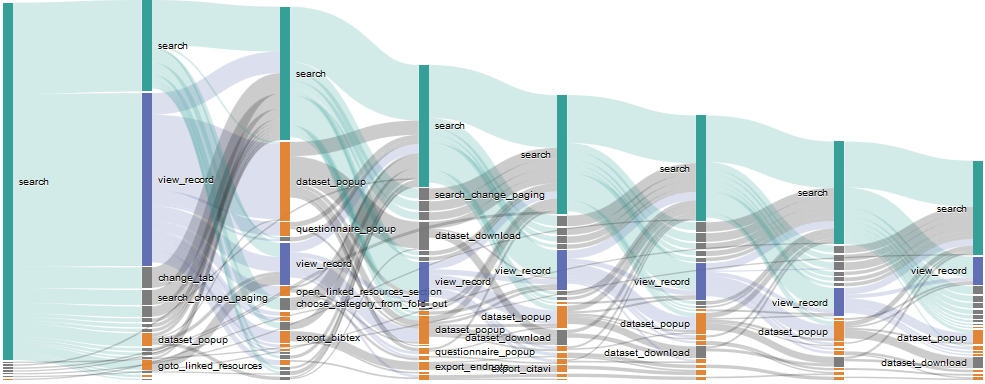}
   \caption{Datasets}
   \label{fig:Ng2}
\end{subfigure}	
   \caption{First eight interaction transitions for publication and dataset search. The interactions are color-coded: green accounts for searching, blue for record view, orange for download (i.e. an implicit relevance signals) and grey for other interactions. 
   Implicit relevance signals indicate a higher degree of relevance suggested by an interaction such as ``export citation'' immediately after a search.    A high resolution diagram is presented in Appendix.
   }
   \label{fig:sankey}
\end{figure}

\vspace{-2mm}
In this section, we study differences between dataset and document search on the basis of interaction sequences. We present this using \emph{Sankey} diagrams in Figure~\ref{fig:sankey}. The diagrams represent the transitions of the first eight interactions of users when searching for publications (Figure~\ref{fig:Ng1}) and datasets (Figure~\ref{fig:Ng2})\footnote{A high-resolution figure is available at: \url{https://arxiv.org/abs/2006.02770}}. In the ISS, it is possible to switch between object types (e.g. from searching for research data to publication search). Hence, we extracted only those sessions from the log having a focus either on publications or on datasets without switching the type in between. Each logged interaction is associated with an action label which describes the type of action a user has performed (``view record'', ``search'' etc.).  An in-depth explanation of this analysis technique can be found in \cite{hienert2016usefulness}.
\vspace{-1.5em}

The analysis of the interaction sequence (see Figure~\ref{fig:sankey}) shows no substantial differences between datasets and publications in terms of interaction paths. 
For both types, the most frequent interactions after an initial search (green) were either a \em view\_record \em (blue) or another \em search\em. 
Differences, however, can be found in two aspects: $a)$ the frequency of consequent searches (green) is higher for publications; $b)$ the number of implicit relevance signals is notably higher for dataset search. 
One can observe that a large fraction of dataset searches contain interactions related to the download of a record which is especially visible in the third interaction for datasets.  Further query reformulations are less frequent for dataset searches (flow into green from any other). 
A possible explanation for this can be that a major portion of dataset searches appear to be known-item based. 
This observation is in line with our earlier observations on session diversity analysis (see Section \ref{diversity}). 

\section{Conclusion and Future Work}
\vspace{-2mm}
In this study, we presented an analysis of search logs from an integrated search system containing both, documents and datasets as repositories.
In contrast to a similar study~\cite{kacprzak-WWW2018}, we experimented with real-life queries issued by social scientists with a defined information need.
Further, we argue that the reported analysis is more factual in accordance with the observations made in~\cite{jansen2006search}. From our study, we observe that the queries addressing a publication are more frequent and less repetitive in comparison to dataset searches. 
Also, the average number of record views during dataset search is substantially lower compared to publication searches.
In terms of segregating search intents between a dataset and a publication search, we note that there are barely any distinctive features to characterize a query.
As part of future work, we would like to utilise the session information to personalise retrieval sessions which can further be used to construct a specialised recommender system for dataset retrieval.
\vspace{-3mm}
\subsection*{Acknowledgement}
This work was funded by DFG under grant MA 3964/10-1, the ``Establishing Contextual Dataset Retrieval - transferring concepts from document to dataset retrieval" (ConDATA) project, \url{http://bit.ly/Condata}.

\bibliographystyle{splncs04}
\bibliography{references} 

\newpage
\section*{Appendix}
%
%
\begin{figure}[H]
 \begin{sideways}
  \begin{minipage}{17.5cm}
   \includegraphics[width=0.9\linewidth]{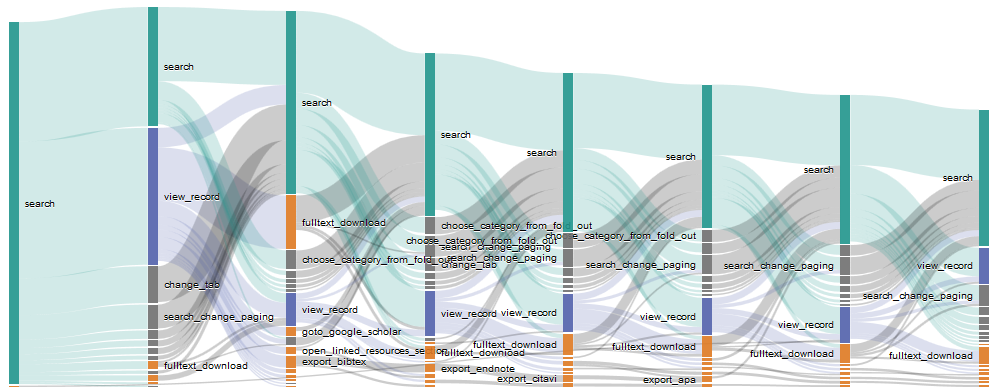}
   \caption{Publications}
   \includegraphics[width=0.9\linewidth]{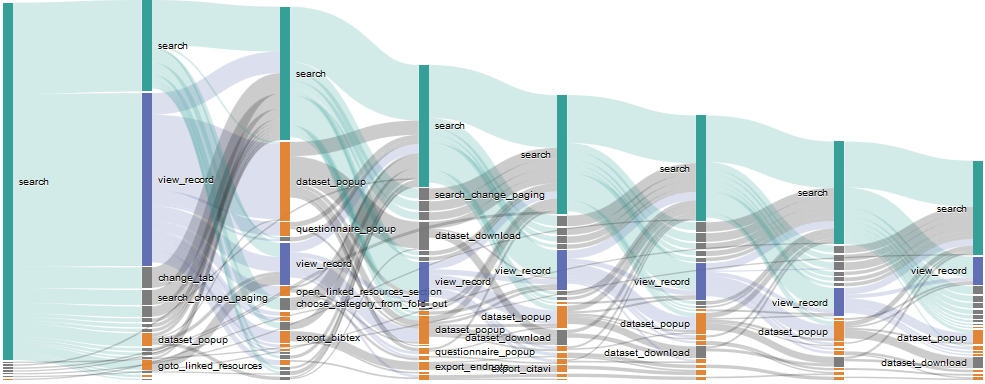}
   \caption{Dataset}
  \end{minipage}
 \end{sideways}
 \centering
  \caption{First eight interaction transitions for publication and dataset search. The interactions are color coded: green accounts for searching, blue for record view, orange for download (i.e. an implicit relevance signals) and grey for other interactions. 
  Implicit relevance signals indicate those interactions with an immediate view record action after search which is an indirect indication of relevance.}
\end{figure}
\end{document}